\documentclass[submission, Proceedings]{SciPost}

\hypersetup{
    colorlinks,
    linkcolor={red!50!black},
    citecolor={blue!50!black},
    urlcolor={blue!80!black}
}

\usepackage[bitstream-charter]{mathdesign}
\DeclareSymbolFont{usualmathcal}{OMS}{cmsy}{m}{n}
\DeclareSymbolFontAlphabet{\mathcal}{usualmathcal}
\begin{document}

\begin{center}{\Large \textbf{
On connection of the cosmic-ray coplanarity of most energetic particles with the collider long-range near-side "ridge" effect\\
}}\end{center}

\begin{center}
R.A. Mukhamedshin\textsuperscript{$\star$}
\end{center}

\begin{center}
Institute for Nuclear Research of Russian Academy of Sciences, Moscow 117312 Russia
\\
* rauf\_m@mail.ru
\end{center}

\begin{center}
\today
\end{center}


\definecolor{palegray}{gray}{0.95}
\begin{center}
\colorbox{palegray}{
  \begin{tabular}{rr}
  \begin{minipage}{0.1\textwidth}
    \includegraphics[width=30mm]{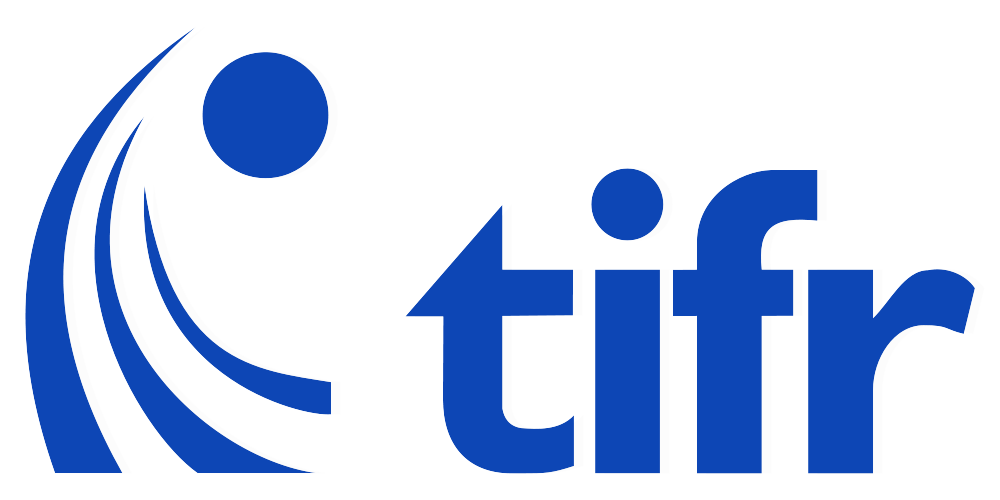}
  \end{minipage}
  &
  \begin{minipage}{0.85\textwidth}
    \begin{center}
    {\it 21st International Symposium on Very High Energy Cosmic Ray Interactions (ISVHECRI 2022)}\\
    {\it Online, 23-27 May 2022} \\
    \doi{10.21468/SciPostPhysProc.?}\\
    \end{center}
  \end{minipage}
\end{tabular}
}
\end{center}

\section*{Abstract}
{\bf
Coplanarity of most energetic subcores of $\gamma$-ray--hadron families observed in cosmic-ray experiments at $E_0 \gtrsim 10^{16}$ eV is explained only with a process of
coplanar generation of most energetic hadrons. Long-range near-side "ridge" effect at $\sqrt{s} = 7$ TeV was found by the CMS Collaboration in two-particle $\Delta\eta -
\Delta\varphi$ correlation functions. The FANSY 2.0 model reproduces both the hadron coplanarity and "ridge" effect. }

\vspace{10pt} \noindent\rule{\textwidth}{1pt} \tableofcontents\thispagestyle{fancy} \noindent\rule{\textwidth}{1pt} \vspace{10pt}

\section{Introduction}
\label{sec:intro}

A tendency to some coplanarity of most energetic subcores of $\gamma$-ray--hadron families (groups of most energetic particles (MEP) in EAS cores) initiated mainly by primary
protons with energies $E_0 \gtrsim 10^{16}$ eV has been observed in X-ray emulsion chamber (XREC) experiments characterized by very high lateral resolution ($\lesssim 1$ mm).

 Five experimental data sets on $\gamma$-ray families with energies $\sum E_\gamma \geq 700$ TeV and $\sum E_\gamma \geq 500$ TeV have been accumulated by the {\em Pamir}
 \cite{Pamir4_1,Kopenetal} and Mt.Canbala \cite{Xue_etal} Collaborations, respectively. Besides, two stratospheric events with extreme energies ($\sum E_\gamma
> 1$ PeV) and coplanarity of MEPs, namely, the {\em Strana} \cite{strana,strana4} and {\em JF2af2} \cite{capdev1}), have been detected.

The coplanarity phenomenon was also studied using the so-called Big Ionization Calorimeter (BIC) \cite{TrudyFIAN1970} (which was a part of a complex facility for studying EAS
at the Tien Shan high-mountain (3340 m a.s.l.) station), with an area of 36 m$^2$  and thickness of five mean free paths for proton interaction. Since the lateral resolution
of the BIC was worse than 0.25 m, main interactions in the selected showers took place at a noticeably higher altitude as compared with the XREC experiments. Preliminary
results showed significant azimuthal effects in large-scale hadronic subcores in EAS cores at a level of six standard deviations above the background at $E_0  =
 0.1 - 10$ PeV  \cite{PavBLPhI1998}. Relevant studies are planned to be continued using a large new ionization calorimeter at the same altitude \cite{ApplSci2023}.

Since the detection of these effects as a result of EAS fluctuations is unlikely  ($\lesssim 10^{-10}$)\cite{MukhJHEP,mukhNP2009,mukhEPJC2009}, they were originally
interpreted as a result of coplanar generation of fragmentation-region MEPs $(x_{ Lab} = E/E_0 \gtrsim 0.01)$ in first interactions of primary protons.

Several theoretical ideas were proposed to explain this phenomenon as a result of (a) conservation of the quark-gluon-string (QGS) angular momentum  \cite{wibig}; (b)
generation of specific leading systems \cite{Royzen,capdev3,Yuldash3,mukhNP99}; (c) evolution of the dimension of space from three to two dimensions \cite{Anchordoquietal}.
 High-$p_{t}^{\, copl}$ values, forming a coplanar plane, are almost necessarily included in the first two concepts \cite{wibig,Royzen,capdev3,Yuldash3,mukhNP99}, while
 $p_{t}$ components directed perpendicular to this plane are traditional. The concept of two-dimensional space \cite{Anchordoquietal} does not require high-$p_{t}^{\, copl}$
 values.

A long-range near-side "ridge" effect in a two-charged-particle  $\Delta \eta - \Delta\varphi$ correlation function, $R_N(\Delta\eta ,\Delta\varphi)$, was  observed by the
CMS Collaboration \cite{ridgeffect} at $|\Delta \eta | \gtrsim 3.0$, $\Delta\varphi \approx 0$. Here $\Delta\varphi$ and $\Delta \eta$ are differences in azimuthal angle
$\varphi$ and pseudorapidity $\eta$, respectively.

The analysis  performed by the CMS Collaboration (and used in this paper) is as follows  \cite{ridgeffect}. First, all events including charged particles with $|\eta | < 2.4$
and
$p_{\textrm{t}} > 0.1$ GeV/c are used to form so-called minimum-bias data set (including all the events) and high-multiplicity one (including collisions with charged-particle
track multiplicity $N^{offline}_{trk} \geq 110$ at $p_{\textrm{t}} > 0.4$ GeV/c). The correlation
function is defined as $
R_N(\Delta\eta ,\Delta\varphi)=\ (\langle N \rangle -1) \left(S_N (\Delta\eta ,\Delta\varphi)/B_N (\Delta\eta ,\Delta\varphi) -1 \right).
$
The signal function
$ S_N(\Delta\eta ,\Delta\varphi) = ({1}/{N^{\rm sign}_{pair}})\, {d^2 N^{\rm sign}}/{d(\Delta\eta) d(\Delta\varphi)} $ and background function $B_N(\Delta\eta ,\Delta\varphi)
= ({1}/{N^{\rm mix} _{pair}})\, {d^2 N^{\rm mix}} /{d(\Delta\eta) d(\Delta\varphi)} $
are determined, respectively,  by counting the number of charged-particle pairs within each event, $N^{\rm sign}_{pair}$, and the number of particle pairs constructed by
randomly selecting two different events (with multiplicities $N_{1}$ and $N_{2}$) and pairing every charged particle within one event with each particle from another event.
Here $N^{\rm mix}_{pair} = N_{1} N_{2}$, $\langle N\rangle $ is the average number of charged particles per event,
 $\Delta\eta$ and $\Delta\varphi$ are always taken to be positive to fill one quadrant of the $\Delta\eta - \Delta\varphi$  histograms (other quadrants are filled by reflection).

\begin{figure}[tb]
\centering
   \includegraphics[width=0.90\textwidth]{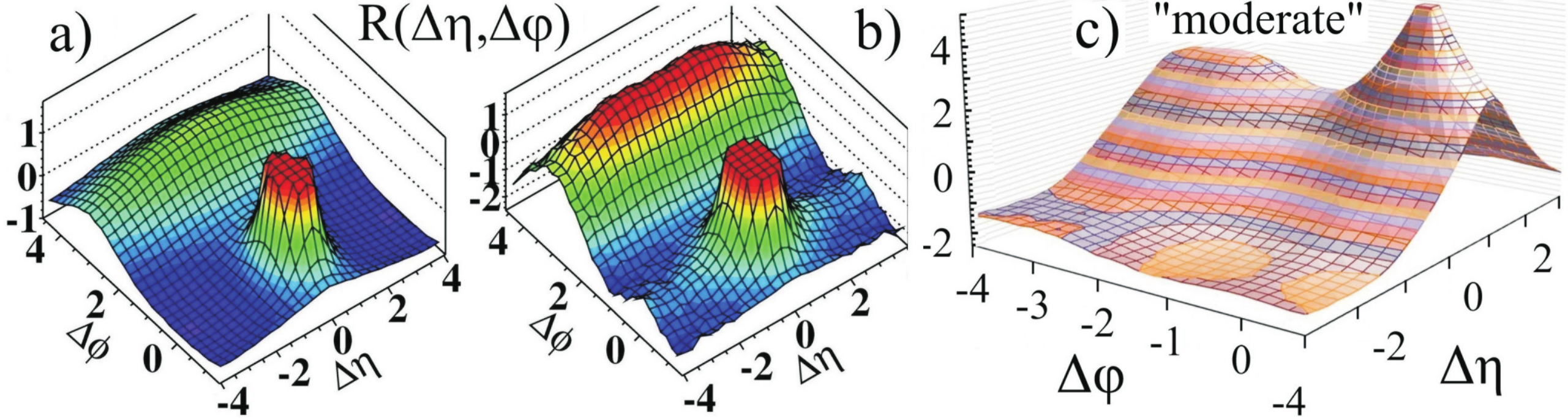}
  \caption{Charged-particle $R (\Delta\eta ,\Delta\varphi)$ functions at $|\Delta\eta | \leq 4$ by the CMS Collaboration for a) minimum-bias and b) high-multiplicity events (Figs. 7b and 7d in Ref. \cite{ridgarXiv1009.4122}); c)  $R (\Delta\eta ,\Delta\varphi)$ function simulated with FANSY 2.0 CPG ("moderate") for minimum bias events.}
  \label{RidgeeffectFANSY}
\end{figure}

In Figs. \ref{RidgeeffectFANSY}a and \ref{RidgeeffectFANSY}b (Figs. 7b and 7d in Ref. \cite{ridgarXiv1009.4122}), experimental $R (\Delta\eta ,\Delta\varphi)$ functions for
minimum-bias and high-multiplicity events are shown. The “ridge” effect is seen in Fig.  \ref{RidgeeffectFANSY}b.

Experiments in cosmic rays and at colliders are carried out under fundamentally different selection criteria. Therefore, it is not possible to directly compare their results.
 However, we can test whether the FANSY 2.0 model\cite{mukhEPJPlus2019}, based on high-$x_{\textrm{F}}$ data and designed to describe the coplanar generation of secondary MEPs, can reproduce any of effects observed in the central kinematic region, in particular, the "ridge" effect.

\section{Coplanarity simulation and results}

The FANSY 2.0  includes conventional  (QGSJ) and coplanar-particle generation (CPG) processes. The QGSJ and CPG versions are similar \cite{mukhEPJC2019} in all
characteristics (except azimuth ones at $\sqrt{s} \gtrsim 2$ TeV), reproduce a lot of  LHC and low-energy data on hadron generation \cite{mukhEPJC2019}, including data on
jets and resonances.
 It should be emphasized that the model was designed specifically for the phenomenological description of the phenomenon of coplanarity of the most energetic EAS subcores, discovered in cosmic-ray experiments, and all important model parameters  were introduced irrespective of the “ridge” effect.

To obtain more detailed conclusions within the FANSY 2.0 model, a large amount of simulations will be carried out, and the results will require separate publications. Until
now, the main goal of this series of works was to show that the predictions of the developed model do not contradict the main results of experiments carried out at the LHC.

The initial concept of MEPs' high-$ p_{\textrm {t}}^{copl}$  coplanarity  \cite{mukhNP2009,mukhEPJC2009} gives a qualitative description of the cosmic-ray phenomenon, but
contradicts LHC data \cite{mukhEPJC2019}. Only a new concept of coplanarity, assuming a reduction in MEPs' transverse-momentum components, directed normal to the coplanarity
plane, makes it possible to reconcile the LHC data and the MEPs' coplanarity \cite {mukhEPJC2019}.
 In this case, the experimental and simulated  $\langle p_\textrm{t}\rangle$ values in all $\eta$ bins do not differ significantly.

The cross section for $pp$ CPG interactions, $\sigma_{\textrm{inel}}^{\rm CPG}(s)$, increases from $\sim 0$  to $42$ mb as $\sqrt{s}$ increases from $\sim 1.25$ to 7 TeV
\cite{mukhEPJC2019}.
 At  $|y|>|y^{\textrm CPG}_{\textrm{thr}}| = |y_{2}| - \Delta_y^{\rm CPG}$ \cite{mukhEPJC2022} (where $y_{2}$ is the rapidity of the second-in-energy particle and only used
to calculate a fluctuating threshold value, $|y^{\textrm CPG}_{\textrm{thr}}|$), the "coplanarization"  algorithm rotates  hadron transverse momentum,
$\overrightarrow{\textbf{p}}_t$, towards the coplanarity plane along the shortest path.
 The azimuthal-angular distribution of MEPs' $\overrightarrow{\textbf{p}}_t$ is
sampled according to the Gaussian distribution (relative to the coplanarity plane) with a standard deviation depending on rapidity as $\sigma^{\rm
CPG}_{\varphi}(y)=\sigma^{\rm CPG}_{\varphi \,0}\cdot (|y_{2}/y|)^\beta$.

Obviously, the most important parameters are those that determine the coplanarity degree, namely, (a) the dispersion of the deviation of the hadron's azimuth angle from the
coplanarity plane, $\sigma^{\rm CPG}_{\varphi \,0}$; (b) the rate of its increase with decreasing rapidity determined by the parameter $\beta$) of the particles, and (c) the
minimum value of MEPs' rapidities, $|y^{\textrm CPG}_{\textrm{thr}}|$ depending on $\Delta_y^{\rm CPG} $). The effective value of $|y^{\textrm CPG}_{\textrm{thr}}|$ can only
be chosen phenomenologically.

In Ref. \cite{mukhEPJC2022} and this paper, similar results were obtained using three slightly  different combinations of the $\Delta_y^{\rm CPG} $, $\sigma^{\rm
CPG}_{\varphi \,0 } $ and $\beta$ parameters (compare Table 1 in both papers), hereinafter referred to as "weak", "moderate" and "strong" versions. The aim was to test the
robustness of the results. The conclusion is in line with expectations, i.e. small changes in parameter values resulted in small changes in the phenomenon magnitude.

\begin{table}
\centering
\caption{\label{effectparamers} Effective parameters of "weak", "moderate", "strong" FANSY 2.0 versions.}
\begin{tabular}{lccc}
\hline\noalign{\smallskip}
   Parameter                                     & "weak" & "moderate" & "strong" \\
\hline\noalign{\smallskip}
$\langle \Delta_y^{\rm CPG}\rangle$              &  3.60  &   4.40     &   4.90   \\
$\langle \sigma^{\rm CPG}_{\varphi \,0} \rangle$ &  0.10  &   0.09     &   0.06   \\
$\langle \beta  \rangle$                         &  1.00  &   0.85     &   0.40   \\

\noalign{\smallskip}\hline \noalign{\smallskip}
\end{tabular}
\end{table}

\begin{figure}[bt]
\centering
   \includegraphics[width=0.90\textwidth]{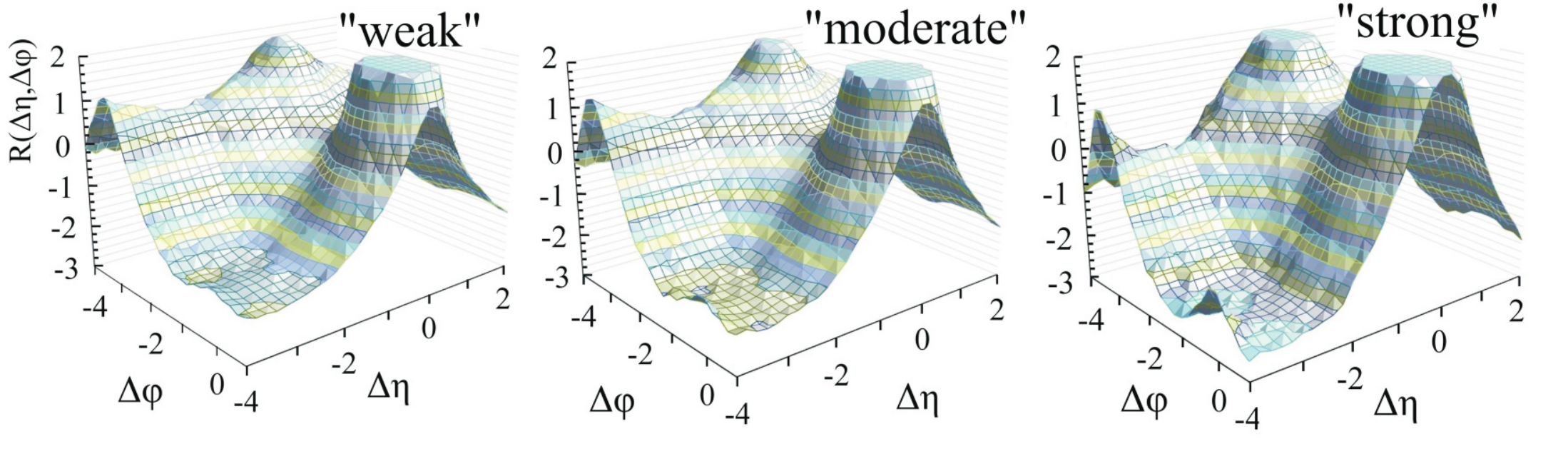}
  \caption{
           The $R (\Delta\eta ,\Delta\varphi)$ functions simulated for high-multiplicity events with "weak", "moderate",  and "strong" FANSY 2.0 CPG versions.}
  \label{eta-phi_R_CPGetaMinWMS}
\end{figure}

The $R (\Delta\eta ,\Delta\varphi)$ function simulated with the FANSY 2.0 QGSJ version for high multiplicity events does not show any "ridge"-like effect as expected (see
Fig. 3 in Ref. \cite{mukhEPJC2022}).
 The $R (\Delta\eta ,\Delta\varphi)$ function simulated with the use of  FANSY 2.0 CPG ("moderate")  for minimum
bias events is shown in Fig. \ref{RidgeeffectFANSY}c, and no significant "ridge"-like effect is also observed in this case.

Figure \ref{eta-phi_R_CPGetaMinWMS} presents $R (\Delta\eta ,\Delta\varphi)$  functions for high-multiplicity events obtained with the use of "weak", "moderate" and "strong"
versions using parameters given in Table \ref{effectparamers}.
 The most strong "ridge" effect is obviously reproduced by the "strong" version, i.e., an increase in the degree of coplanarity of MEPs causes a corresponding increase in the effect in the correlation function.

\begin{figure}[tb]
\centering
    \includegraphics[width=0.90\textwidth]{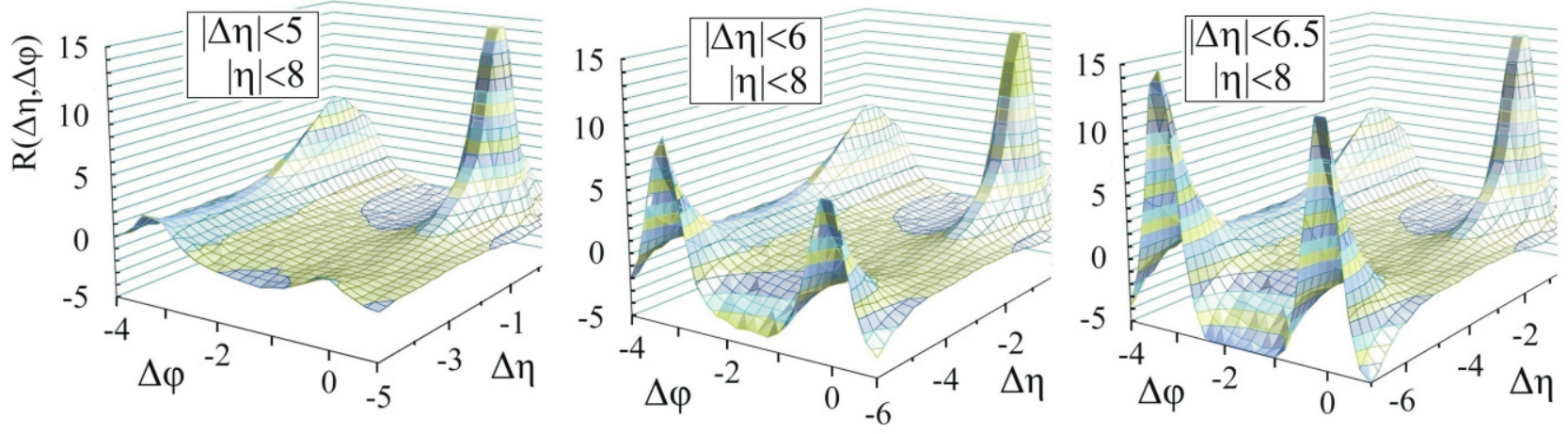}
   \caption{$R (\Delta\eta ,\Delta\varphi)$ distributions simulated by FANSY 2.0 CPG ("moderate") for high-multiplicity events at  $|\Delta\eta | \leq 5$, $|\Delta\eta | \leq 6$, $|\Delta\eta | \leq 6.5$ and $|\eta | \leq 8$.}
   \label{eta-phi_R_CPGeta8_5665}
 \end{figure}

All the above-considered results were obtained with $|\Delta\eta | \leq 4$ and $|\eta | \leq 2.4$. If  the $|\Delta \eta |$ interval could be expanded, then interesting
effects would be observed for high-multiplicity events. Figure \ref{eta-phi_R_CPGeta8_5665} shows the $R(\Delta\eta ,\Delta\varphi)$ function simulated by FANSY 2.0 CPG
("moderate") at $|\Delta\eta | \leq 5$, $|\Delta\eta | \leq 6$, $|\Delta\eta | \leq 6.5$, while $|\eta|$ values of particles under consideration vary in a wider range,
namely, $|\eta | \leq 8.0$. At $|\Delta\eta| > $4, the "ridge" effect will rapidly develops into a brighter phenomenon expressed in the appearance of two peaks close in size
("twin peaks") at $|\Delta\varphi| \approx 0$ and $|\Delta\varphi| \approx \pi$.
 So, Fig. \ref{eta-phi_R_CPGeta8_5665} shows that an expansion of the $|\Delta\eta |$ range can inform us on features of the MEP coplanarity region. If the peak growth is found to
stop at $|\Delta\eta | \sim 5$, and amplitude of peaks  begin to decrease with a further increase in $|\Delta\eta |$, this means that the "ridge" effect is not associated
with the process of coplanar generation of MEPs.

Thus, there is a qualitative agreement between the experimental and simulated $R (\Delta\eta ,\Delta\varphi)$  functions in the region of the "ridge" effect, while there are
differences between the functions at $|\Delta \varphi| \approx \pi$.  It is the CPG process that creates some growth of the simulated $R (\Delta\eta ,\Delta\varphi)$
functions at $|\Delta\varphi| \approx \pi$  and $|\Delta\eta | \gtrsim 3$.

Within the FANSY 2.0 CPG model, the coplanarity planes have the same azimuthal orientation in the opposite hemispheres. The "twin peaks" effect at $|\Delta\eta|\gtrsim 4$  is
determined mainly by hadrons from different hemispheres, both at $|\Delta\varphi|\sim 0$ and  $|\Delta\varphi|\sim \pi$. The coplanarization algorithm is a very primitive
procedure. So, one cannot rule out, for example, different orientations of the coplanarity planes in the opposite hemispheres (or even more complex cases)that will entail
corresponding changes in $R (\Delta\eta ,\Delta\varphi)$ functions.

In any case, the fact that FANSY 2.0 (being an \textit{ad hoc} model based on high-$x_{\textrm{F}}$ data) reproduces both the coplanarity effect and long-range near-side
"ridge" effect with a slight change in parameters, may indicate a qualitative relationship between these phenomena.

\section{Search for coplanarity signatures at the LHC}

Simulations show that CPG signatures can appear in the range $5.25 < \eta < $6.5 \cite{mukhEPJC2019}, so the CMS-CASTOR very-forward calorimeter experiment \cite{CASTOR},
focused on this $|\eta |$ interval, seems promising. For a simplified assessment of CASTOR capabilities, let us assume that it consists of 16 radially arranged segments
divided in half by a vertical slit. If the pseudorapidities of particles (which do not fall into the slit) are in the range $5.25 < \eta < 6.5$, particles are considered to
be registered, and their energies are considered to be "measured". 

Only interactions with the total "measured" energy $\sum E_i > 1$ TeV are considered. Here $E_i$ is the energy value "measured" in the $i$-th segment ($1 \leq i \leq 16$).
 In addition, the number of segments $N_s$, in each of which the "measured" energy $E_i > E_{i\, min} = 100$ GeV, must be equal to or greater than two. The segment with the
  maximum energy release, $E_{max}$, gets the first number, i.e. $E_1=E_{max}$. The remaining segments are numbered clockwise in ascending order. Finally, let us define
  variables as follows: $E_{\rm copl}=E_1 + E_9$; $E_{tr}=E_5+E_{13}$;  $\varepsilon_{\rm copl} = E_{\rm copl}/(E_{\rm copl}+E_{tr})$. Obviously, at $\varepsilon_{\rm copl}=1$, the degree of event coplanarity is maximum.

\begin{figure}[bt]
\centering
   \includegraphics[width=0.50\textwidth]{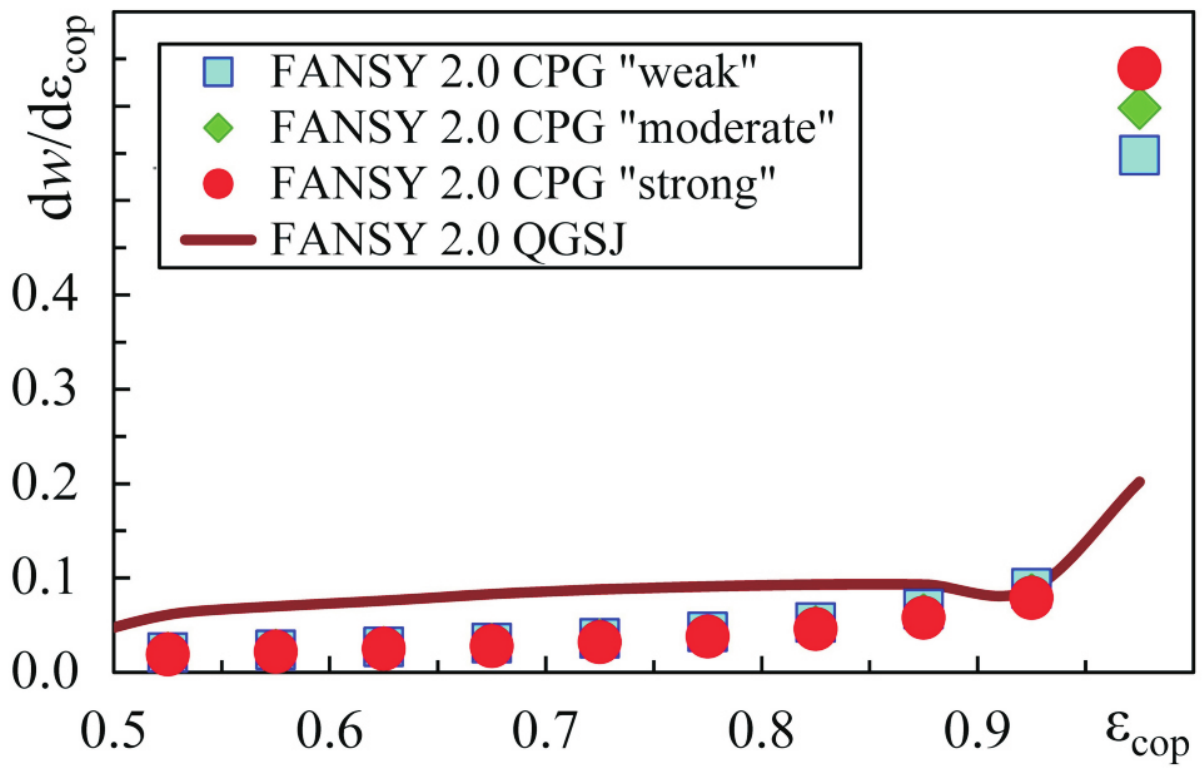}
  \caption{Probability $d\omega/d\varepsilon_{\rm copl}$ distributions predicted by "weak", "moderate",  and "strong" FANSY 2.0 CPG versions for the CASTOR detector. }
  \label{figCASTOR}
\end{figure}

Figure \ref{figCASTOR} shows $d\omega/d\varepsilon_{\rm copl}$ probability distributions  obtained using the QGSJ version (line), as well as the "weak", "moderate" and
"strong" CPG versions. As expected, the "strong" CPG version predicts the highest value of $d\omega/d\varepsilon_{\rm copl}$ at $\varepsilon_{\rm copl} \rightarrow 1$.
  Undoubtedly, a detailed study of the coplanarity phenomenon requires much more accurate simulations.

\section{Conclusion}

The long-range near-side "ridge" effect observed by the CMS Collaboration can be a by-product of the coplanar generation of most energetic particles in hadronic interactions.

A significant "twin peaks" effect is predicted, which appears in $R (\Delta\eta ,\Delta\varphi)$ correlation functions at $|\Delta\eta | > 4$ in high-multiplicity events

Coplanarity signatures could be observed with the CMS-CASTOR very-forward calorimeter.




\nolinenumbers


\begin{thebibliography}{99}

\bibitem{Pamir4_1} A. Borisov et al. (Pamir Collaboration), Proc. 4th ISVHECRI, Beijing  (1986) 4-29
\bibitem{Kopenetal} V.V. Kopenkin et al.,  Phys. Rev. D \textbf{52} {2766} (1995)

\bibitem {Xue_etal} L. Xue  et al.,  Proc. 26th Int. Cosmic Ray Conf., Salt Lake City (1999)  {\bf 1} 127

\bibitem{strana}  A.V. Apanasenko et al., Proc. 15th  Int. Cosmic Ray Conf., Plovdiv (1977)  {\bf 7} 220
\bibitem{strana4} A.K. Managadze et al., Physics of Atomic Nuclei \textbf{70} \textbf{1} (2007) 184

\bibitem{capdev1} J.N. Capdevielle. J.Phys. G \textbf{14} (1988) {503}

\bibitem{TrudyFIAN1970} T.P. Amineva, T.G. Glavach, V.S. Aseikin  et al., Trudy FIAN \textbf{46} (1970) 157 (in Russian).
\bibitem{PavBLPhI1998} V.P. Pavlyuchenko. Bull. Lebedev Phys. Inst., 1998, No. 6, 1 – 7 (in Russian).
\bibitem{ApplSci2023}  R.A. Mukhamedshin, T. Sadykov, A. Serikkanov et al.,  Appl. Sci. 2023, 13(4), 2507.

\bibitem{MukhJHEP} R.A. Mukhamedshin, JHEP  \textbf{05} (2005) 049
\bibitem{mukhNP2009} R.A. Mukhamedshin, Nucl. Phys. B (Proc. Suppl.)  \textbf{196C} (2009) 98
\bibitem{mukhEPJC2009} R.A. Mukhamedshin, Eur. Phys. J. C  \textbf{60} (2009) 345

\bibitem{wibig} T. Wibig, hep-ph/0003230
\bibitem{Royzen} I.I. Royzen. Mod. Phys. Lett. A \textbf{9}  (1994) no.38 3517
\bibitem{capdev3} J.N. Capdevielle, Nucl. Phys. B (Proc. Suppl.) \textbf{175 - 176}  (2008) 137
\bibitem{Yuldash3} T.S. Yuldashbaev et al., Nuovo Cim. \textbf{24C} (2001) 569
\bibitem{mukhNP99} R.A. Mukhamedshin, Nucl. Phys. B (Proc. Suppl.) \textbf{75A}  (1999) {141}

\bibitem{Anchordoquietal} L.A. Anchordoqui, De Chang Dai, H. Goldberg et al., Phys. Rev. D \textbf{83} (2011) 114046

\bibitem{ridgeffect} The CMS Collaboration, JHEP \textbf{09} (2010) 091

\bibitem{mukhEPJC2022} R. A. Mukhamedshin. Eur. Phys. J. C (2022) 82:155
\bibitem{mukhEPJPlus2019} R.A. Mukhamedshin. Eur. Phys. J. Plus (2019) 134: 584

\bibitem{ridgarXiv1009.4122} The CMS Collaboration, arXiv:1009.4122v1 [hep-ex] 21 Sep 2010

\bibitem{mukhEPJC2019} R.A. Mukhamedshin. Eur. Phys. J. C (2019) 79: 441
\bibitem{CASTOR} The CMS collaboration. JINST \textbf{16} (2021) P02010
\bibitem{mukhBRAS2021} R.A. Mukhamedshin. Bull. Russ. Acad. Sci. Phys. 85, 402–404 (2021)




\end{thebibliography}
\end{document}